\documentclass[aps,prd,twocolumn,superscriptaddress,nofootinbib,showkeys,showpacs]{revtex4}

\usepackage{epsfig}
\usepackage{amsmath}
\usepackage{amssymb}

\newcommand{\etal}{{\em et al}}

\begin{document}

\preprint{FERMILAB-PUB-09-624-A}

%
%
\title{Closing the Window on Strongly Interacting Dark Matter with IceCube}

\author{Ivone F.~M.~Albuquerque}\affiliation{Center for Particle Astrophysics, Fermi National Accelerator Laboratory, 
Batavia, IL, 60510. USA}
\affiliation{Instituto de F\'isica, Universidade de S\~ao Paulo, S\~ao Paulo, Brazil}
\author{Carlos P\'erez~de los Heros}\affiliation{Department of Physics and Astronomy. Uppsala University. Uppsala. Sweden}

\date{\today}
 
\begin{abstract}
 We use the recent results on dark matter searches of the 22-string IceCube detector to 
probe the remaining allowed window for strongly interacting dark matter 
in the mass range 10$^4<$m$_\text{\tiny X}<$10$^{15}$ GeV. 
We calculate the expected signal in the 22-string IceCube detector from the annihilation of 
such particles captured in the Sun and compare it to the detected background. 
As a result, the remaining allowed region in the  mass versus cross section 
parameter space is ruled out. We also show the expected sensitivity of the 
complete IceCube detector with 86 strings.
\end{abstract}


\maketitle

\section{\label{sec:Intro} Introduction}
 The search for dark matter is currently one of the most active fields of research in 
experimental astroparticle physics. Common candidates are weakly interacting 
massive particles (WIMPs), with only weak and gravitational interactions with normal matter, 
and which encompass a variety of particle types in the mass range from a few tens of GeV to 
a few hundred TeV, where the upper limit is based on theoretical arguments to preserve unitarity~\cite{Griest:90a}. 
Among WIMP  candidates are the lightest neutralino arising in the Minimal Supersymmetric Extension of 
the Standard Model (MSSM), or the lightest Kaluza-Klein mode  in models of Universal Extra 
Dimensions (UED)~\cite{Hooper:07a}. These candidates are thermal relics from the Big Bang, and since they 
are stable, assumed to be able to contribute to the dark matter content in the halos of galaxies.\par

 In a different scenario, it has been shown that super-massive particles can be produced in the early Universe 
and account for dark matter, independently of their interaction strength with normal matter. In order to avoid 
the mass limit imposed by unitarity constraints, one can model non--thermal production of super-massive 
particles~\cite{Chung:98a,Chung:99a}. The large mass will prevent the particle to ever get into thermal equilibrium 
with the primordial plasma. Super-massive dark matter candidates were coined wimpzillas~\cite{Chung:98b} if assumed to 
interact weakly with matter and simpzillas~\cite{Albuquerque:01a} if the interaction is strong. The production mechanism 
discussed in~\cite{Chung:98a,Chung:99a} favours particles with masses of the order of the inflaton mass ($\sim$10$^{12}$~GeV), 
but particles of masses as low as a few hundred GeV can also be produced, normally denoted as strongly interacting massive 
particles (SIMPs). Although SIMPs are in general taken to be in the mass range $\sim$10$^4\lesssim$ m$_\text{\tiny X}\lesssim$10$^8$ GeV, 
and simpzillas, with masses m$_\text{\tiny X}>$10$^8$ GeV, we will generically call them either simpzillas. \par

\begin{figure}[t]
\centering\epsfig{file=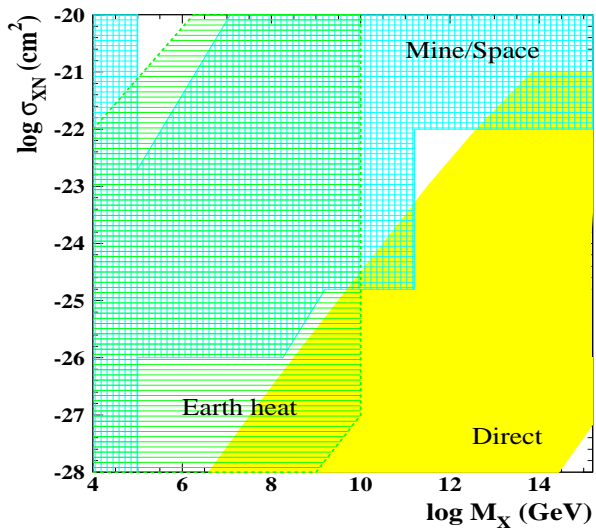,height=0.8\linewidth,width=0.9\linewidth}
\caption{Excluded region at 90\% C.L. in the simpzilla mass versus cross section
parameter space. The region labeled ``Direct'' (solid yellow region) was
excluded based on direct dark matter detection~\cite{Albuquerque:03a}; the ``Earth heat''
region (green striped) is excluded based on the Earth's heat flow 
\cite{Mack:07a} and the blue square--hatched ``Mine/Space'' region is based on
many experiments, underground and space-borne~\cite{McGuire:01a,Starkman:90a}.} 
\label{fig:pexc}
\end{figure}

Searches for simpzillas have been carried out in a variety of ways in the past. Figure~\ref{fig:pexc}
shows the current exclusion region in the mass versus simpzilla--nucleon cross section parameter space. The ``Mine/Space'' 
region (blue square hatched) was ruled out by many different experiments underground or space-borne~\cite{McGuire:01a,Starkman:90a}~\footnote{Note 
that part of the exclusion region shown in~\cite{Starkman:90a} and reproduced in \cite{Mack:07a} was 
corrected in~\cite{McGuire:01a}}. More recently, a search for simpzillas using results from direct detection 
experiments excluded most of the remaining  parameter space for high masses (solid yellow, labeled ``Direct'')~\cite{Albuquerque:03a}. 
A recent analysis  further constrained the allowed region by using the Earth heat flow to set constrains on the 
annihilation of captured dark matter in the Earth's center (green striped, labelled ``Earth heat'')~\cite{Mack:07a}. 
The blank triangular regions in Figure~\ref{fig:pexc} remained unprobed until this work.\par

 Whether thermally produced or not, stable dark matter candidates can gravitationally accumulate as a halo in our galaxy, 
becoming bound in orbits in the solar system. Further energy losses can occur in elastic interactions with matter in large 
celestial bodies, like the Sun. If they lose enough energy they can fall below the escape velocity of the object, accumulating 
in its center and annihilating therein~\cite{Press:85a,Gaisser:86a,Gould:88a}. Neutrinos of energies {\it{O}}(100 GeV) can be 
produced from the decays of the annihilation products. This allows for an 'indirect search' for dark matter with neutrino telescopes, 
based on a search for an excess neutrino flux  from the direction of the Sun over the known atmospheric neutrino background. 
Predictions of simpzilla event rates and neutrino telescopes sensitivity  have been anticipated in~\cite{Albuquerque:02a}.\par

In this paper we explore the capabilities of IceCube to detect high-energy neutrinos from simpzilla annihilations in the Sun. 
 The baseline IceCube geometry consists of 80 strings with 60 digital optical modules each, instrumenting 1 km$^3$ 
of ice at depths between 1450~m--2450~m near the geographic South Pole~\cite{IceCube:04a}. Recently, an addition of six 
strings forming a denser core in the middle of the IceCube array has been proposed in order to lower the neutrino energy threshold 
of the detector to about  10~GeV, the so--called ``DeepCore'' array~\cite{IceCube:DeepCore}. Although the IceCube geometry has been  
optimized to detect ultra-high energy ($>$TeV) neutrinos from potential cosmic sources with sub-degree angular resolution, its 
 angular response is still adequate at {\it{O}}(100 GeV) to perform directional searches for dark matter. IceCube has recently published 
results on the searches for WIMPs arising in the MSSM~\cite{IceCube:WIMP} or the simplest UED model~\cite{IceCube:KK} using 22 strings 
deployed in the 2006/2007 Antarctic season. In this work we use this 22-string detector (hereafter referred to as IceCube-22) which took 
data during the 2007 austral winter. We set a limit in the simpzilla mass, m$_\text{\tiny X}$, versus simpzilla-nucleon  cross 
section, $\sigma_{\text{\tiny{XN}}}$, phase space using the fact that IceCube-22 has not detected any neutrino excess from the direction 
of the Sun. We also estimate the sensitivity of the full IceCube detector in its expected final configuration with 86 strings 
(hereafter referred to as IceCube-86).\par

\begin{figure}[t]
\centering\epsfig{file=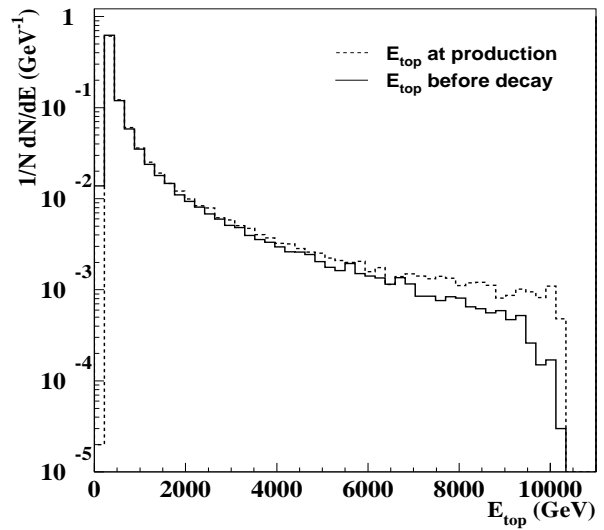,height=0.8\linewidth,width=0.9\linewidth}
\caption{Energy distribution of the top quarks produced in simpzilla annihilations at production point (dashed line) and before 
decaying (full line). The figure shows that initial-state gluon radiation does not play an important role in degrading the 
energy of the tops before they decay.}
\label{fig:E_t}
\end{figure}

\section{\label{sec:spectrum} Neutrino spectrum from simpzilla annihilations} 
In order to estimate the neutrino rate from simpzilla annihilations in the Sun, 
we use the capture rate, as well as the full-flavour neutrino flux and energy spectrum at the core of the Sun 
as determined in~\cite{Albuquerque:01a}. We then simulate the neutrino propagation to the Earth, including energy losses and oscillation effects.\par 

 The capture rate, $\Gamma_c$, depends on the mass of the simpzilla, m$_\text{\tiny X}$, and the strength of the interaction of simpzillas with 
nucleons, $\sigma_{\text{\tiny{XN}}}$. The capture efficiency can be described by the parameter 
$q=\frac{m_\text{\tiny X}}{m_\text{\tiny N}\,n_\text{\tiny N}\,\sigma_{\text{\tiny{XN}}} R_\odot}$, where  m$_\text{\tiny N}$ is the average 
nucleon mass, n$_\text{\tiny N}$ is the number density of nucleons in the Sun and R$_\odot$ is the radius of the Sun. For large cross 
sections, q $\leq$ 1, and simpzillas lose enough energy in their passage through the Sun to be efficiently captured.  
The capture rate (in s$^{-1}$) is then given by 
\begin{equation}
\begin{split}
\Gamma_c\, &=\,10^{17} (1+y^2)\left(\frac{10^{12}\, \text{GeV}}{m_\text{\tiny X}}\right)\,\times \\
 &\times \left( \frac{\text{u}_{th}}{240\, \text{km}\, \text{s}^{-1}}\right) \,\left( \frac{R_\odot}{7\times 10^{10} \text{cm}}\right)^2 \\
\end{split}
\label{eq:Gamma1_C}
\end{equation}

On the other hand, if q$>$1, capture is determined by the relative velocity, and only low velocity simpzillas are trapped. In such case, 
the capture rate is 
\begin{equation}
\begin{split}
\Gamma_c &= 10^{17} [1+y^2- e^{-x^2}(1+y^2+x^2) ]\left(\frac{10^{12}\, \text{GeV}}{m_\text{\tiny X}}\right)\,\times \\ 
& \times \left( \frac{\text{u}_{th}}{240\, \text{km}\, \text{s}^{-1}}\right) \,\left( \frac{R_\odot}{7\times 10^{10} \text{cm}}\right)^2 \\
\end{split}
\label{eq:Gamma2_C}
\end{equation}

where in the above expressions 
\begin{equation}
y\,=\, 2.5 \left(\frac{v_\odot}{600\, \text{km}\, \text{s}^{-1}}\right)\,\left( \frac{\text{u}_{th}}{240\, \text{km}\, \text{s}^{-1}}\right)^{-1}\;\text{and}\;x=\frac{y}{\sqrt{q-1}}
\label{eq:xy}
\end{equation}
and u$_{th}$ is the speed of the Sun in the galaxy and $v_\odot$ is the Sun escape velocity.  
Note that if $\sigma_{\text{\tiny{XN}}}$ goes to zero, then the capture rate goes to zero through its dependency on $x$, as it 
is expected. \par 
The simpzilla capture rate is large enough to reach equilibrium with the annihilation rate, $\Gamma_A$,
in the lifetime of the solar system, 
and therefore we can assume~$\Gamma_A~\!\!=~\!\!\Gamma_c~\!\!/2$. \\
\begin{figure}[t]
\centering\epsfig{file=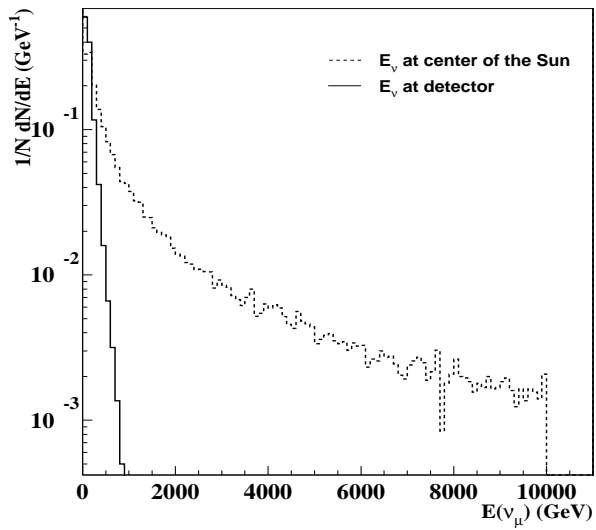,height=0.8\linewidth,width=0.9\linewidth}
\caption{Muon-neutrino energy distribution from top decays at the center of the Sun (dashed) and at the IceCube location (full line), 
after energy losses and oscillations have been taken into account.}
\label{fig:nuflux}
\end{figure}
 The neutrino spectrum from simpzilla annihilations is modeled as producing
a pair of quarks or gluons, 
which will fragment into high multiplicity hadronic jets. Long-lived light and charmed hadrons will lose energy  
through interactions in the dense solar interior before decaying, producing low energy neutrinos. On the other hand, due to its short lifetime, 
$\sim 10^{-25}$ s, the top quark will not have time to hadronize or lose energy through interaction in the medium before decaying 
into W$b$.  The W then decays into $l\nu_l$, with similar probability into each flavour $l=e, \mu, \tau$. Annihilations into top quarks is therefore 
a promising channel to produce high energetic neutrinos detectable by neutrino telescopes. Bottom quarks are also a potential source of 
high energy neutrinos~\cite{Albuquerque:01a}. However in this work we only consider neutrinos produced from the top decay chain, which makes our 
results conservative. Note  that although the initially available energy for each top is very high (the mass m$_\text{\tiny{X}}$ of the simpzilla), 
the available phase space is shared among many produced particles (2.8$\times$10$^5\sqrt{m_\text{\tiny{X/12}}}\,$ tops are produced per 
annihilation~\cite{Albuquerque:01a}, where $m_\text{\tiny{X/12}}$ is the mass of the simpzilla in units of 10$^{12}$~GeV), so the average energy 
per produced particle is of {\it{O}} (TeV). The main consequence of this is that initial-state gluon radiation can be neglected, not playing an 
important role in further degrading the energy of the tops before they decay. This is illustrated in Figure~\ref{fig:E_t}, which shows the energy 
distribution of the tops at the production point (dashed line) and just before decaying (full line), obtained with a full \texttt{PYTHIA}~\cite{PYTHIA} simulation. \par
 Taking into account the above considerations, the energy distribution of the neutrinos produced from top decays at the center of the Sun 
can be parametrized as

\begin{equation}
\frac{dN}{dE_{\nu}}\propto 
 \frac{E_{\nu}+\text{m}_{\mbox{\tiny W}}}{\sqrt{(E_{\nu}+\text{m}_t)[(E_{\nu}+\text{m}_t)^2-\text{m}_t^2][(E_{\nu}+\text{m}_{\mbox{\tiny W}})^2-\text{m}_{\mbox{\tiny W}}^2]}}\\
\label{eq:parametrization}
\end{equation}
where m$_{\mbox{\tiny W}}$ and m$_t$ are the W boson and top quark masses respectively, and we have omitted a constant normalization factor. \par
 The energy dependence of the neutrino flux at the detector (Earth) will be modified by neutrino energy losses in their way 
out of the Sun and by oscillations. In order to include these effects we use the publicly available  
Wimpsim code~\cite{Edsjo:00a}, which we have modified so it takes the neutrino spectrum from equation~(\ref{eq:parametrization}). We use as 
input the three flavours with their correct relative intensity from the top decay chain. Figure~\ref{fig:nuflux} shows the injection spectrum of 
muon neutrinos at the Sun from simpzilla annihilations according to equation~(\ref{eq:parametrization}) (dashed line), and the resulting spectrum at 
the Earth (full line). The injection spectrum has been truncated at 10 TeV for display purposes. As mentioned earlier,  the initial energy available in 
the annihilation is shared among an enormous amount of annihilation products and, furthermore, the energy of the resulting neutrinos is 
further reduced by energy losses in the Sun. The spectrum at the detector is therefore not very different from those of lighter dark matter 
candidates, like MSSM neutralinos or Kaluza-Klein modes. This allows us to use IceCube-22 background determination found in both WIMP and 
Kaluza-Klein dark matter analysis \cite{IceCube:WIMP,IceCube:KK}. Since IceCube is mainly sensitive to muon neutrinos through the detection of muon 
tracks, we only use the muon-neutrino flux at the detector for our calculations in what follows.

\begin{figure}[t]
\centering\epsfig{file=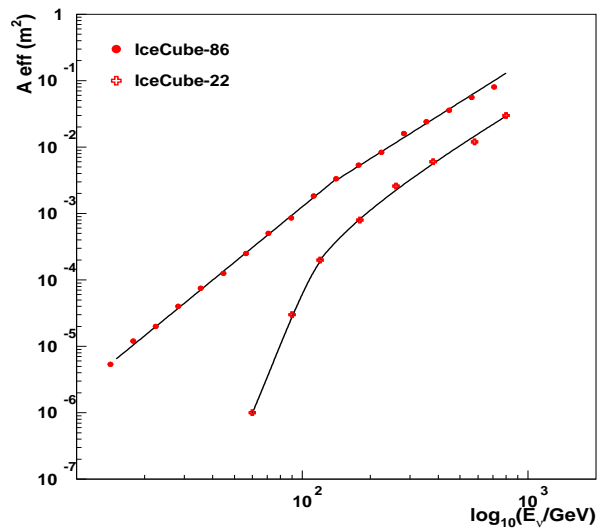,height=0.8\linewidth,width=0.9\linewidth}
\caption{Neutrino effective area of IceCube, both the 22-string detector and an estimation for the planned 86-string final configuration, as 
a function of neutrino energy.}
\label{fig:IceCube_Aeff}
\end{figure}

\section{\label{sec:IceCube22} Results from IceCube-22}
The number of signal events, N$_s$, from simpzilla annihilations 
in the Sun predicted in a neutrino telescope of effective area A$_{eff}$ during an exposure time T, is given by
\begin{equation}
N_s(m_\text{\tiny X}, \sigma_{\text{\tiny{XN}}})= N_t \cdot BR_{\mbox{\tiny W}} \cdot \Gamma_A(m_\text{\tiny X},\sigma_{\text{\tiny{XN}}})\cdot T \cdot \int \frac{dN_{\nu}}{dE}\,A_{eff}\,dE
\label{eq:n_events}
\end{equation}
where $dN_{\nu}/dE$ is the muon neutrino flux at the detector, $N_t$ is the number of tops per 
annihilation, $BR_{\mbox{\tiny W}}$ is the branching ratio of W decaying into neutrinos (0.326) and $\Gamma_A$ is the annihilation rate in 
the center of the Sun. The effective area is a measure of the efficiency of the detector and includes the neutrino-nucleon interaction 
probability, the energy loss of the produced muon from the interaction point to the detector and the detector trigger and analysis efficiency. \par
 
\begin{figure}[t]
\centering\epsfig{file=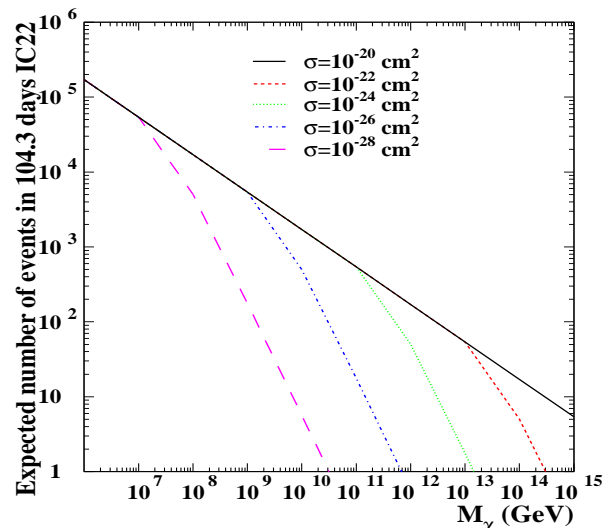,height=0.8\linewidth,width=0.9\linewidth}
\caption{Expected number of events in the IceCube-22 detector, in 104.3
days of lifetime. The rate is determined by our simulations and is shown
for five different values of the cross section as labeled.}
\label{fig:evr}
\end{figure}

In order to compare our expected signal with IceCube-22 observations, we use the neutrino effective area of IceCube-22 published in~\cite{IceCube:KK}, 
and shown in Figure~\ref{fig:IceCube_Aeff}. In 104.3 days of live time IceCube-22 has detected 13 events from the direction of the Sun, while 
the expected number of atmospheric neutrinos is 18.5~\cite{IceCube:KK}.  The expected signal strength for the given live time is 
shown in Figure~\ref{fig:evr}. We have assumed a 15\% uncertainty in the expectation of the atmospheric neutrino flux at the relevant energies 
(E$_{\nu}\geq$50 GeV) in order to calculate the 90\% CL upper limit on a possible signal, $\mu^{90}_s$, following the prescription in~\cite{Rolke:04a}. 
We obtain $\mu^{90}_s$ = 3.1. A scan in the  (m$_\text{\tiny X}$, $\sigma_{\text{\tiny{XN}}}$) parameter space, selecting models that predict less events 
than $\mu^{90}_s$  leads to the 90\% CL exclusion region shown in Figure~\ref{fig:exclusion} (read cross--hatched area). The results disfavour a large 
region of the parameter space, and we discuss their significance in section~\ref{sec:conclusions}.

\section{\label{sec:IceCube86} Sensitivity of the completed IceCube-86 detector}
The sensitivity of the completed IceCube detector can be estimated from the expected number of background events  
(due to atmospheric neutrinos) and the detector effective area.  An estimation of the expected effective area of IceCube-86 at trigger level 
has been given in~\cite{IceCube:DeepCore}. Previous analyses of IceCube and AMANDA show that the effective area is reduced between a factor 
of 5 to 10 between trigger level and the final analysis level, due to the  cuts that are applied to clean the data sets 
and select high quality tracks. We have taken a conservative approach here and we have reduced the trigger effective area 
given in~\cite{IceCube:DeepCore} by a factor between 15 for low energies down to 5 at the higher energies, and used that as 
our estimate of where the effective area of IceCube-86 at a final analysis level might be. The curve used is 
shown as the upper curve in Figure~\ref{fig:IceCube_Aeff}. \par

\begin{figure}[t]
\centering\epsfig{file=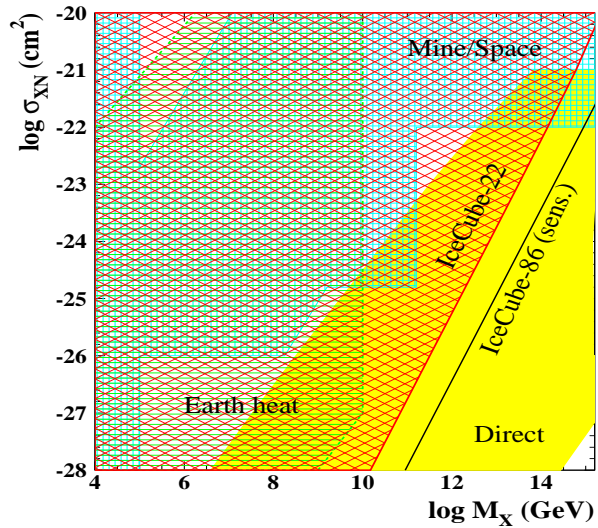,height=0.8\linewidth,width=0.9\linewidth}
\caption{Disfavoured region in the (m$_\text{\tiny X}$, $\sigma_{\text{\tiny{XN}}}$) parameter space at 90\% confidence level 
from IceCube-22 (red cross--hatched area), overlayed with results from previous experiments: direct searches~\cite{Albuquerque:03a} (solid yellow area), 
space-borne detectors~\cite{McGuire:01a} (blue square--hatched area) and Earth heat flow analysis~\cite{Mack:07a} (green striped area).  The black line represents the 
1 year expected sensitivity of the completed IceCube detector.
}
\label{fig:exclusion}
\end{figure}

We have calculated the number of atmospheric neutrino events in IceCube using the measured atmospheric neutrino flux 
by AMANDA~\cite{IceCube:atmflux}. We  parametrized the curve in Figure~10 in~\cite{IceCube:atmflux} 
as dN$^{atm}_{\nu}$/dEd$\Omega$ = A E$^{-\gamma}$ GeV$^{-1}$ cm$^{-2}$ s$^{-1}$ sr$^{-1}$, with A=0.11407 and $\gamma$ = 3.142. 
The expected number of atmospheric neutrino events in IceCube-86 from the direction of the Sun in a exposure time T is then given by 
$N_{\nu}~=~T ~\cdot~\int~dN^{atm}_{\nu}/dEd\Omega\,A_{eff}\,dE~d\Omega$, where the angular integration is performed over 
the 0.5$^\circ$ that the Sun subtends in the sky.  We have assumed no contamination of misreconstructed 
atmospheric muons in the final sample, and therefore we use only the 3.4 atmospheric neutrino events 
from the direction of the Sun predicted by the integral above in 365 days live time.  As in the previous section, we have 
used a 15\% uncertainty  on the overall normalization of the atmospheric neutrino flux, as also indicated by the 
AMANDA measurement, and calculated the sensitivity including this uncertainty. The 90\% CL exclusion region from 
IceCube-86 is shown in Figure~\ref{fig:exclusion} as the black line. It agrees well with the predictions made in~\cite{Albuquerque:02a}. 

\section{\label{sec:conclusions} Conclusions} 
 We have used recently published results from the 22-string IceCube detector on WIMP dark matter 
searches~\cite{IceCube:WIMP,IceCube:KK} to extend the search to strongly interacting 
dark matter candidates in the mass range  $\sim$10$^4\lesssim$ m$_\text{\tiny X}\lesssim$10$^{15}$ GeV. 
The fact that the IceCube results are compatible with the expected atmospheric neutrino background allows us 
to set restrictive limits on the mass versus nucleon cross section parameter space.  \par
A note on the complementarity of direct and indirect searches is due at this point. Generally, the total cross section with nucleons,  
$\sigma_{\text{\tiny{XN}}}$, can have at least two contributions, a spin-independent (or scalar) component and a spin-dependent (or axial-vector) 
component, the relative strength of each  depending on the exact assumptions made on the structure and interaction of the simpzilla. 
 There is no detailed modelling in the literature of the way that simpzillas interact with matter, 
 other than the interaction is assumed to be ``strong'', as opposed to weak as in the case of WIMPs. 
However for experimental searches such distinction is important because different experiments probe different components of the cross section. 
The results from the experiments shown in Figure~\ref{fig:pexc} refer mainly to the spin-independent component of the simpzilla-nucleon cross 
section, due to the nature of the targets involved or, in the case of the Earth heat flow analysis, due to the fact that the capture of dark matter candidates in the Earth 
proceeds mainly through scattering on spinless nuclei (Fe, Si, O, Mg). On the other hand the IceCube results are sensitive to the spin-dependent 
cross section since the Sun is the target for capture, which is essentially a proton target. \par
 In view of the previous considerations, the current results from IceCube provide the most restrictive limits so far on the spin-dependent cross section of simpzillas. The fact that they overlap in the (m$_\text{\tiny X}$, $\sigma_{\text{\tiny{XN}}}$) parameter  
space with previous exclusion regions from experiments which probe mainly the spin-independent cross section, practically rules out 
simpzillas as dark matter. It would require an unreasonably narrow fine tuning of the mass and the structure of the cross section  
to propose a viable candidate that would evade all the current limits.  \par
 In conclusion, strongly interacting  heavy relics from the early Universe can only account for the dark matter if they have masses above
$10^{15}$~GeV. As these ultra heavy masses are not currently favored by any model, the window for simpzillas as the only 
component of dark matter can be considered closed.\\

\noindent{\small{\bf Acknowledgments:}}
We thank T. Sj\"{o}strand for his kind help with \texttt{Pythia}. IA was partially funded by the U.S. Department 
of Energy under contract number DE-AC02-07CH11359 and the Brazilian National Counsel for Scientific Research (CNPq)

\end{document}